\documentclass[10pt,a4paper]{iopart}\eqnobysec
\usepackage{graphicx}
\usepackage{cite}

\newcommand{\adj}{^{\dagger}}
\newcommand{\D}{{\mathrm{d}}}
\newcommand{\Exp}[1]{\rme\power{#1}}
\newcommand{\EXP}[1]{\exp\left(#1\right)}
\newcommand{\power}[1]{^{\mbox{\footnotesize$#1$}}}
\newcommand{\rewop}[1]{_{\mbox{\footnotesize$#1$}}}
\newcommand{\proj}[1]{|#1\rangle\langle#1|}
\newcommand{\ds}{\displaystyle}
\newcommand{\clst}[1]{\left\lfloor #1 \right\rceil}
\newcommand{\half}{\frac{1}{2}}
\newcommand{\thalf}{{\textstyle\frac{1}{2}}}
\newcommand{\ket}[1]{\left| #1 \right\rangle}

\newcommand{\bra}[1]{\left\langle #1 \right|}

\newcommand{\braket}[2]{\bigl\langle #1\big| #2 \bigr\rangle}
\newcommand{\expect}[1]{\bigl\langle #1 \bigr\rangle}
\newcommand{\trc}[1]{\tr\left\{ #1 \right\}}
\newcommand{\lesssim}{\mathrel{\rule{2pt}{0pt}%
\raisebox{1.8pt}{\makebox[0pt][c]{$<$}}%
\raisebox{-2pt}{\makebox[0pt][c]{\small$\sim$}}\rule{2pt}{0pt}}}
\newcommand{\sinc}{\mathop{\mathrm{sinc}}}
\newcommand{\snc}{\mathop{\mathrm{snc}}}
\newcommand{\sgn}{\mathop{\mathrm{sgn}}}

\newcommand{\eqref}[1]{(\ref{eq:#1})}

\newcommand{\myPR}[3]{#2 \textit{Phys.\ Rev.\/} \textbf{#1} #3}
\newcommand{\myPRL}[3]{#2 \textit{Phys.\ Rev.\ Lett.\/} \textbf{#1} #3}

\newcommand{\JPhysA}[3]{#2 \textit{J.\ Phys.\ A: Math.\ Gen.\/} \textbf{#1} #3}
\newcommand{\myAPNY}[3]{#2 \textit{Ann.\ Phys.\ (NY)\/} \textbf{#1} #3}

\begin{document}

\title{Periodic and discrete Zak bases}

\author{Berthold-Georg Englert\dag, 
Kean Loon Lee\dag,
Ady Mann\ddag\  
and Michael Revzen\ddag}

\address{\dag 
Department of Physics, National University of Singapore, 
Singapore 117542, Singapore}

\address{\ddag
Department of Physics, Technion---Israel Institute of Technology, 
Haifa 32000, Israel}

\ead{phyebg@nus.edu.sg}

\begin{abstract}
Weyl's unitary operators for displacement in position and momentum commute
with one another if the product of the elementary displacements equals Planck's
constant. 
Then, their common eigenstates constitute the Zak basis, with each state
specified by two phase parameters.
Accordingly, the transformation function from the position basis to the Zak
basis maps the Hilbert space on the line onto the Hilbert space on the torus.
This mapping is one-to-one provided that the Zak basis states are periodic
functions of their phase parameters, but then the mapping cannot be continuous
on the whole torus. 
\newline
With the periodicity of the Zak basis enforced, the basis has a double Fourier
series. 
The Fourier coefficients identify a discrete basis which complements 
the periodic Zak basis to form a pair of mutually unbiased bases.
The discrete basis states are the common eigenstates
of the two complementary partners to the two unitary displacement operators.
These partner operators are of angular momentum type, with integer
eigenvalues, and generate the fundamental rotations of the torus.
Conversely, the displacement operators are the ladder operators for their
partners.
\newline
For each consistent phase convention for the periodic Zak basis, and thus for
the line-onto-torus mapping, there is a corresponding discrete Zak basis and a
corresponding pair of partner operators. 
Examples of particular interest are the conventions that give a
continuous mapping in one phase parameter or are symmetric in both phase
parameters. 
The latter emphasizes the Heisenberg--Weyl symmetry between position and
momentum.  
\newline
We discuss briefly the relation between the Zak bases and Aharonov's modular
operators. 
Finally, as an application of the Zak operators for the torus, 
we mention how they can be used to associate with the single degree 
of freedom of the line a pair of genuine qubits that are potentially 
entangled.  
\end{abstract}


\submitto{\JPA}

\maketitle

\section{Introduction}
A convenient choice of variables may render simple a complex
physical problem. 
Joshua Zak introduced \cite{Zak1,Zak2} what he termed ``the
$kq$ representation'' as a particularly convenient mode of
description for electrons in solids \cite{Zak3,Zak4}. 
Zak's work was based on Weyl's \cite{Weyl} presentation of unitary
displacement operators for coordinates and momenta (obtained by
exponentiating both the momentum and the coordinate operators) and
noting that for some particular choice of parameters these displacement
operators commute. 
Weyl's work \cite{Weyl} led to several studies that dealt with both
mathematical and physical 
representations wherein the spatial coordinates are defined modulo
some conveniently selected length $a$ and the momenta modulo
$2\pi/a$ (in units for which $\hbar=1$). 
However, it was Zak \cite{Zak1,Zak2} who was the first to recognize the wide
context of the approach and to study it in a systematic way \cite{Janssen1}. 

The Zak transform of a state may be regarded as a mixed position-momentum 
(or time-frequency) representation of the state. 
The transform enabled
Zak \cite{Zak3,Zak4} to reduce the problem of an electron in a periodic
potential in a constant magnetic field to a form that made possible
studies of further physical applications --- to digital data
transmission, for instance, as is considered in \cite{Janssen2}. 

Reference \cite{Janssen1} is a thorough review of the Zak transform
and its applications up to the year 1988;
for recent applications, see \cite{Zak3a,Zak3b,Zak3c} for example.
In the more recent literature the representation finds its widest use in
signal processing where time plays the role of the spatial coordinate
and frequency that of the momentum; see
\cite{Janssen2,Gabor1,Gabor2} 
and references therein.

In the present work, 
which we regard as a contribution to our understanding of quantum kinematics, 
we consider the Zak transform as a mapping of the Hilbert
space on a line to the Hilbert space on a torus. 
This allows us to study, in section~\ref{sec:stage}, 
periodic and discrete Zak bases which are mutually unbiased 
\cite{Wootters1,Wootters2}. 
In section~\ref{sec:periodicWFs} we use some
freedom in the permissible definition of the phase of the periodic
wave functions in the Zak basis to present and discuss three
convenient phase choices. 
Section~\ref{sec:discreteWFs} is devoted to the study of the
discrete Zak basis; we display and discuss the Wigner functions
\cite{Wigner1} (adopting the conventions of \cite{Wigner2}; see the footnote
at \eqref{lmWigner} below)
that are associated with the wave functions in these bases. 
In section~\ref{sec:OperRels} the position and momentum operators are given
as differential operators in the Zak representation of 
section~\ref{sec:periodicWFs}.
Section~\ref{sec:ModOps} deals with the relation of the Zak bases and
operators to the modular variables introduced by Aharonov \emph{et al.\ } 
in Ref.~\cite{Aharonov}. 
A possible application for the Zak bases in quantum information theory
is hinted at in section~\ref{sec:TorQbits}. 
We close with a Summary.

\section{Setting the stage}\label{sec:stage}
We consider a single continuous quantum degree of freedom, the motion along
the $x$ axis, for which the position operator $X$ and the momentum operator
$P$ constitute the fundamental complementary pair of hermitian dynamical 
variables.
They obey the Heisenberg--Born commutation relation
$\bigl[X,P\bigr]=\rmi\hbar$, and we adopt the usual normalization conventions
for their complete sets of eigenbras and eigenkets.

We denote Weyl's unitary operator for the momentum shift $P\to P-p_0$ by $U$
and that for the position shift $X\to X-x_0$ by $V$,
\begin{equation}
  \label{eq:UVdef}
  U=\Exp{-\rmi p_0 X/\hbar}\,,\qquad V=\Exp{\rmi x_0 P/\hbar}\,.
\end{equation}
When both are acting on an operator, a function of $X$ and $P$, their order
is irrelevant,
\begin{eqnarray}
  \label{eq:UVonF}
  f(X,P)\to f(X-x_0,P-p_0)&=U\adj V\adj f(X,P)UV\nonumber\\
                          &=V\adj U\adj f(X,P)VU\,.
\end{eqnarray}
But the order does matter for the transformation of bras and kets, as
illustrated by 
\begin{eqnarray}
  \label{eq:UVonBra}
  \bra{x}UV=\Exp{-\rmi p_0x/\hbar}\bra{x}V
           =\Exp{-\rmi p_0x/\hbar}\bra{x+x_0}\,,
\nonumber\\
\bra{x}VU=\bra{x+x_0}U=\Exp{-\rmi p_0(x+x_0)/\hbar}\bra{x+x_0}\,,
\end{eqnarray}
unless the product $p_0x_0$ is an integer multiple of Planck's constant
$h=2\pi\hbar$.
Following Zak \cite{Zak1,Zak2}, we therefore choose the two displacements such
that 
\begin{equation}
  \label{eq:p0x0}
  p_0x_0=2\pi\hbar\,,\qquad UV=VU\,.
\end{equation}
The product $p_0x_0$ then equals the phase space area that is roughly
associated with one quantum state, and it is fitting to regard $p_0$ and $x_0$
as corresponding atomic units for momentum and position. 
This is emphasized, for example, by the relations
\begin{eqnarray}
  U=\Exp{-2\pi\rmi  X/x_0}\,,\qquad V=\Exp{2\pi\rmi P/p_0}\,,\nonumber\\
2\pi\rmi\bigl[P/p_0,X/x_0\bigr]=1\,.
\end{eqnarray}

Now that $U$ and $V$ commute, they have common eigenstates, which we label by
the real phases $\alpha$ and $\beta$ that are associated with the unit-modulus
eigenvalues,
\begin{eqnarray}
  \label{eq:alpha-betaDef}
  U\ket{\alpha,\beta}=\ket{\alpha,\beta}\Exp{-\rmi\alpha}\,,\nonumber\\
  V\ket{\alpha,\beta}=\ket{\alpha,\beta}\Exp{\rmi\beta}\,.
\end{eqnarray}
Since the characterizing eigenvalues are $2\pi$-periodic in $\alpha$
and $\beta$, we require that the eigenstates are periodic as well,
\begin{equation}
  \label{eq:periodicKets}
  \ket{\alpha,\beta}=\ket{\alpha+2\pi,\beta}=\ket{\alpha,\beta+2\pi}\,.
\end{equation}
In the completeness relations, then,
\begin{equation}
  \label{eq:compl}
  \int_{(2\pi)}\!\frac{\D\alpha}{2\pi}\int_{(2\pi)}\!\frac{\D\beta}{2\pi}\,
  \proj{\alpha,\beta}=1\,,
\end{equation}
the integration is over any $2\pi$ interval, and the wave functions of the Zak
representation, $\psi(\alpha,\beta)=\braket{\alpha,\beta}{\ }$, are functions
on the torus, not on the square.

In the orthonormality relation
\begin{equation}
  \label{eq:orthoZak}
  \braket{\alpha,\beta}{\alpha',\beta'}
 =(2\pi)^2\delta^{(2\pi)}(\alpha-\alpha')\delta^{(2\pi)}(\beta-\beta')
\end{equation}
we meet the periodic delta function
\begin{equation}
  \label{eq:delta2pi}
  \delta^{(2\pi)}(\varphi)=\sum_{k=-\infty}^{\infty}\delta(\varphi-2\pi k)
=\frac{1}{2\pi}\sum_{k=-\infty}^{\infty}\Exp{\rmi k\varphi}\,.
\end{equation}
It appears also in the explicit construction of the projector on an
${\alpha,\beta}$ state in terms of the unitary shift operators,
\begin{eqnarray}
  \label{eq:proj}
  \proj{\alpha,\beta}&=\sum_{j,k=-\infty}^{\infty}
      \left(\Exp{\rmi\alpha}U\right)^j\left(\Exp{-\rmi\beta}V\right)^k
\nonumber\\
&=(2\pi)^2\delta^{(2\pi)}(p_0X/\hbar-\alpha)
\delta^{(2\pi)}(x_0P/\hbar-\beta)\,.
\end{eqnarray}

The Fourier coefficient kets $\ket{l,m}$ that are defined by the double Fourier
series of $\ket{\alpha,\beta}$,
\begin{equation}
  \label{eq:Fourier}
  \ket{\alpha,\beta}
  =\sum_{l,m=-\infty}^{\infty}\ket{l,m}\Exp{-\rmi(l\alpha-m\beta)}\,,
\end{equation}
form the discrete Zak basis that is normalized in accordance with
\begin{equation}
  \label{eq:discZak}
  \braket{l,m}{l',m'}=\delta_{mm'}\delta_{ll'}\,,\qquad
\sum_{l,m=-\infty}^{\infty}\proj{l,m}=1\,.
\end{equation}
Upon observing that
\begin{equation}
  \label{eq:MUB}
  \braket{\alpha,\beta}{l,m}=\Exp{\rmi(l\alpha-m\beta)}\,,\qquad
\bigl|\braket{\alpha,\beta}{l,m}\bigr|=1\,,
\end{equation}
we note that the periodic and the discrete Zak bases are mutually unbiased.

We regard the discrete basis states $\ket{l,m}$ as common eigenstates of two
integer operators,
\begin{equation}
  \label{eq:L12Def}
L\ket{l,m}=\ket{l,m}l\,,\qquad M\ket{l,m}=\ket{l,m}m\,.
\end{equation}
Their action on the continuous Zak states amounts to differentiation,
\begin{equation}
  \label{eq:LsDiff}
  \bra{\alpha,\beta}L
=\frac{1}{\rmi}\frac{\partial}{\partial\alpha}\bra{\alpha,\beta}\,,\qquad  
  \bra{\alpha,\beta}M
=\rmi\frac{\partial}{\partial\beta}\bra{\alpha,\beta}\,,
\end{equation}
which implies that they are the generators of the fundamental rotations of the
torus,
\begin{equation}
  \label{eq:rotateTorus}
  \bra{\alpha,\beta}\Exp{\rmi(\alpha'L-\beta'M)}
 =\bra{\alpha+\alpha',\beta+\beta'}\,.
\end{equation}
Conversely, the unitary displacement operators are ladder operators for the
discrete Zak basis,
\begin{equation}
  \label{eq:ladderUV}
  U^{l'}V^{m'}\ket{l,m}=\ket{l-l',m-m'}\,.
\end{equation}
The commutation relations
\begin{eqnarray}
  \label{eq:CRs}
  \bigl[U,V\bigr]=0\,,\qquad&\bigl[U,L\bigr]=U\,,\qquad&\bigl[U,M\bigr]=0\,,
\nonumber\\
  \bigl[V,M\bigr]=V\,,\qquad&\bigl[M,L\bigr]=0\,,\qquad&\bigl[V,L\bigr]=0
\end{eqnarray}
reveal that the pairs $(U,L)$ and $(V,M)$ are the dynamical variables of
two \emph{independent} degrees of freedom of azimuth--angular-momentum
type.  
These are, of course, the two degrees of freedom of the torus.

\section{Wave functions: Periodic Zak basis}\label{sec:periodicWFs}
As a consequence of the eigenket equations \eqref{alpha-betaDef},
the position wave function $\braket{x}{\alpha,\beta}$ of the periodic Zak
state, which is the transformation function between the position
representation and the periodic Zak representation, obeys the functional
equations 
\begin{eqnarray}
  \braket{x}{\alpha,\beta}
  &=\Exp{\rmi(\alpha-p_0x/\hbar)}\braket{x}{\alpha,\beta}\nonumber\\
  &=\Exp{-\rmi\beta}\braket{x+x_0}{\alpha,\beta}\,.
\end{eqnarray}
The corresponding equations for the momentum wave function are
\begin{eqnarray}
  \braket{p}{\alpha,\beta}
  &=\Exp{\rmi\alpha}\braket{p+p_0}{\alpha,\beta}\nonumber\\
  &=\Exp{-\rmi(\beta-x_0p/\hbar)}\braket{p}{\alpha,\beta}\,.
\end{eqnarray}
Their general solution is 
\begin{eqnarray}
  \label{eq:cont-wf}
  \braket{x}{\alpha,\beta}&=
\frac{2\pi}{\sqrt{x_0}}\,\chi(\alpha,\beta)\,
\Exp{-\rmi\frac{\alpha\beta}{4\pi}}\Exp{\rmi\beta x/x_0}\,
\delta^{(2\pi)}(p_0x/\hbar-\alpha)\,,
\nonumber\\
  \braket{p}{\alpha,\beta}&=
\frac{2\pi}{\sqrt{p_0}}\,\chi(\alpha,\beta)\,
\Exp{\rmi\frac{\alpha\beta}{4\pi}}\Exp{-\rmi\alpha p/p_0}\,
\delta^{(2\pi)}(x_0p/\hbar-\beta)\,,
\end{eqnarray}
where $\chi(\alpha,\beta)$ must be of unit modulus,
\begin{equation}
  \label{eq:modChi}
 \bigl|\chi(\alpha,\beta)\bigr|=1\,,  
\end{equation}
for consistency with the normalization \eqref{compl},
and the periodicity of \eqref{periodicKets} requires that $\chi(\alpha,\beta)$ 
obeys
\begin{equation}
  \label{eq:periodChi}
  \chi(\alpha,\beta)=\chi(\alpha+2\pi,\beta)\Exp{-\rmi\beta/2}
                    =\Exp{\rmi\alpha/2}\chi(\alpha,\beta+2\pi)\,.
\end{equation}
Any $\chi(\alpha,\beta)$ permitted by these constraints is an
acceptable phase convention for the periodic Zak bases.

Equation \eqref{periodChi} implies that
\begin{equation}
  \label{eq:shiftChi}
  \chi(\alpha,\beta)=(-1)\power{ab}\,\Exp{\rmi(a\beta-\alpha b)/2}\,
\chi(\alpha-2\pi a,\beta-2\pi b)
\end{equation}
if $a,b$ are integers,
so that $\chi(\alpha,\beta)$ is specified by stating its values within a
standard ${2\pi\times 2\pi}$ square.
We find the following three choices for $\chi(\alpha,\beta)$ particularly
natural and interesting:
\begin{eqnarray}
\label{eq:threeChoices}
\mbox{(a)}\ \chi(\alpha,\beta)=
\Exp{\rmi\alpha\left(\frac{\beta}{4\pi}-\clst{\frac{\beta}{2\pi}}\right)}\,,
\nonumber\\
\mbox{(b)}\ \chi(\alpha,\beta)=
\Exp{-\rmi\beta\left(\frac{\alpha}{4\pi}-\clst{\frac{\alpha}{2\pi}}\right)}\,,
\\\nonumber
\mbox{(c)}\ \chi(\alpha,\beta)=
(-1)\power{\clst{\frac{\alpha}{2\pi}}\clst{\frac{\beta}{2\pi}}}
\Exp{\rmi\left(\clst{\frac{\alpha}{2\pi}}\frac{\beta}{2}
-\frac{\alpha}{2}\clst{\frac{\beta}{2\pi}}\right)}\,,
\end{eqnarray}
where
\begin{equation}
  \label{eq:closest}
  \clst{z}
   =\int\rewop{0}\power{z}\D\zeta\,
           \sum_{k=-\infty}^{\infty}\delta(\zeta-k-\thalf)
   =z-\frac{1}{\pi}\arctan\bigl(\tan(\pi z)\bigr)
\end{equation}
denotes the integer closest to $z$.
The $\chi(\alpha,\beta)$ of choice (a) is continuous in
$\alpha$ but discontinuous in $\beta$; that of choice (b) is discontinuous in
$\alpha$ and continuous in $\beta$; and the one resulting from choice (c) is
symmetric in $\alpha$ and $\beta$ in the sense of
$\chi(\alpha,\beta)=\chi(-\beta,\alpha)=\chi(\beta,\alpha)^*$ and
discontinuous in both.  

As these examples demonstrate, discontinuities are a generic feature of the
wave functions we are dealing with here, and of other functions derived from
them. 
It will be necessary on various occasions to assign values to such functions at
their discontinuities.
We employ the convention to assign half the sum
of the one-sided limits, for which
\begin{equation}
  \label{eq:signum}
\eqalign{\mathrm{sgn}(x)=\left\{
    \begin{array}{rl}
     +1 & \mbox{for $x>0$,}\\
      0 & \mbox{for $x=0$,}\\
     -1 & \mbox{for $x<0$,}\\
    \end{array}\right.
\\ 
\bigl[\mathrm{sgn}(x)\bigr]^2=1\quad\mbox{for all $x$,}
\\
  \Exp{\rmi\varphi\,\mathrm{sgn}(x)}=\left\{
    \begin{array}{ll}
     \Exp{\rmi\varphi} & \mbox{for $x>0$,}\\
      \cos{\varphi} & \mbox{for $x=0$,}\\
     \Exp{-\rmi\varphi} & \mbox{for $x<0$,}\\
    \end{array}\right.}
\end{equation}
are illustrative examples.
As an application, and for future reference, we note that
\begin{eqnarray}
  \label{eq:diffChi}
\fl  \chi(\alpha,\beta)^*\frac{1}{\rmi}\frac{\partial}{\partial\alpha}
\chi(\alpha,\beta)=\left\{\begin{array}{@{}l}
\mbox{(a)}\ 
\ds\frac{\beta}{4\pi}-\clst{\frac{\beta}{2\pi}}\,,\\[2ex]
\mbox{(b)}\ 
\ds-\frac{\beta}{4\pi}+\delta^{(2\pi)}(\alpha-\pi)\sin\beta\,,\\[2ex]
\mbox{(c)}\ 
\ds-\half\clst{\frac{\beta}{2\pi}}
+\delta^{(2\pi)}(\alpha-\pi)(-1)\power{\clst{\frac{\beta}{2\pi}}}
\sin\frac{\beta}{2}\,;
  \end{array}\right.
\nonumber \\[-1ex] \qquad\\[-1ex] \nonumber
\fl  \chi(\alpha,\beta)^*\rmi\frac{\partial}{\partial\beta}
\chi(\alpha,\beta)=\left\{\begin{array}{@{}l}
\mbox{(a)}\
\ds-\frac{\alpha}{4\pi}+\delta^{(2\pi)}(\beta-\pi)\sin\alpha\,,\\[2ex]
\mbox{(b)}\ 
\ds\frac{\alpha}{4\pi}-\clst{\frac{\alpha}{2\pi}}\,,\\[2ex]
\mbox{(c)}\ 
\ds-\half\clst{\frac{\alpha}{2\pi}}
+\delta^{(2\pi)}(\beta-\pi)(-1)\power{\clst{\frac{\alpha}{2\pi}}}
\sin\frac{\alpha}{2}\,;
  \end{array}\right.
\end{eqnarray}
for the respective choices in \eqref{threeChoices}.

Further, as an important consequence of \eqref{periodChi}, 
we have the two Fourier series
\begin{eqnarray}
  \label{eq:mulam2chi}
  \chi(\alpha,\beta)
  &=\Exp{-\rmi\frac{\alpha\beta}{4\pi}}
    \sum_{m=-\infty}^\infty\lambda(\alpha-2\pi m)\Exp{\rmi m\beta}
\nonumber\\
  &=\Exp{\rmi\frac{\alpha\beta}{4\pi}}
    \sum_{l=-\infty}^\infty\Exp{-\rmi l\alpha}\mu(\beta-2\pi l)
\end{eqnarray}
with
\begin{eqnarray}
  \label{eq:chi2mulam}
  \lambda(\alpha)&=\int_{(2\pi)}\frac{\D\beta}{2\pi}\,
                  \Exp{\rmi\frac{\alpha\beta}{4\pi}}\,\chi(\alpha,\beta)\,,
\nonumber\\
  \mu(\beta)&=\int_{(2\pi)}\frac{\D\alpha}{2\pi}\,
                  \Exp{-\rmi\frac{\alpha\beta}{4\pi}}\,\chi(\alpha,\beta)\,,
\end{eqnarray}
where the integrands are $2\pi$ periodic functions of the respective
integration variable. 
Either one of the single-argument functions $\lambda(\alpha)$ or $\mu(\beta)$
introduced here, 
which are related to each other by Fourier transformation,
\begin{eqnarray}
  \label{eq:mulam}
  \lambda(\alpha)&=\int_{-\infty}^{\infty}\frac{\D\beta}{2\pi}\,
               \Exp{\rmi\frac{\alpha\beta}{2\pi}}\mu(\beta)\,,
\qquad
\mu(\beta)&=\int_{-\infty}^{\infty}\frac{\D\alpha}{2\pi}\,
                \Exp{-\rmi\frac{\alpha\beta}{2\pi}}\lambda(\alpha)\,,
\end{eqnarray}
can thus be used for a complete specification of $\chi(\alpha,\beta)$.

The restriction imposed by the normalization \eqref{modChi} can be stated as
\begin{equation}
  \label{eq:orthomu}
  \int_{-\infty}^{\infty}\frac{\D\alpha}{2\pi}
\Bigl[\Exp{\rmi l\alpha}\lambda(\alpha-2\pi m)\Bigr]^*
\Bigl[\Exp{\rmi l'\alpha}\lambda(\alpha-2\pi m')\Bigr]
=\delta_{ll'}\delta_{mm'}
\end{equation}
or equivalently as
\begin{equation}
  \label{eq:completemu}
  \sum_{l,m=-\infty}^{\infty}
\Bigl[\Exp{\rmi l\alpha}\lambda(\alpha-2\pi m)\Bigr]
\Bigl[\Exp{\rmi l\alpha'}\lambda(\alpha'-2\pi m)\Bigr]^*
=2\pi\delta(\alpha-\alpha')
\end{equation}
or by the corresponding equations for $\mu(\beta)$.
Equation \eqref{orthomu} has the appearance of an orthonormality relation,
while \eqref{completemu} looks like a completeness relation,
which is, of course, not accidental and will be clarified in
section~\ref{sec:discreteWFs}.  

\begin{figure}[h]
\centerline{%
\begin{picture}(325,240)(0,0)
\put(5,0){\includegraphics{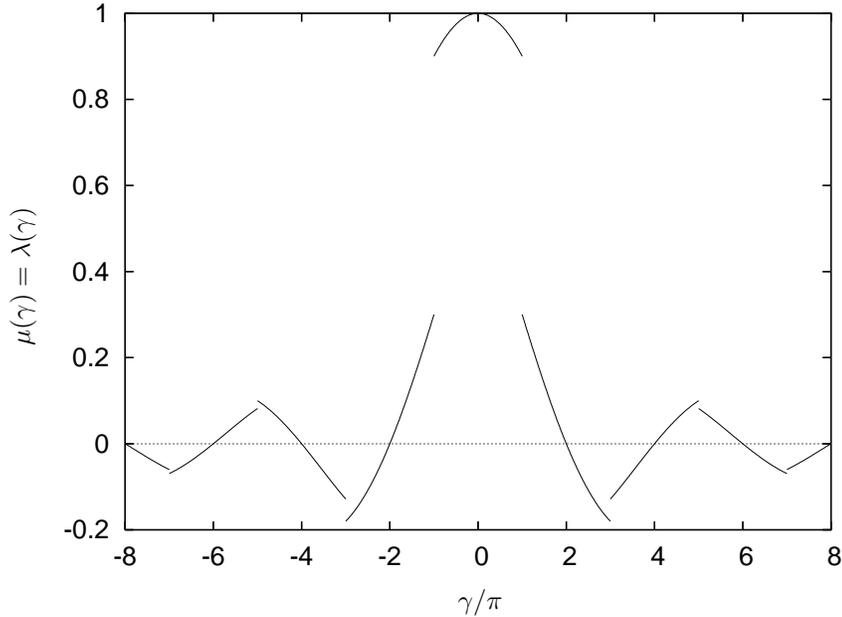}}
\end{picture}
}
\caption{\label{fig:mulam-c}%
The discontinuous function $\lambda(\gamma)=\mu(\gamma)$ of choice (c) in
equation \eqref{3mulam} which is its own Fourier transform.} 
\end{figure}

For the particular choices of \eqref{threeChoices}, we have
\begin{eqnarray}
  \label{eq:3mulam}
  \mbox{(a)}\ & \lambda(\alpha)=\frac{2}{\alpha}\sin\frac{\alpha}{2}
               \equiv\sinc\left(\frac{\alpha}{2}\right)\,,
\nonumber\\
             & \mu(\beta)=\left\{\begin{array}{cc}
               1 & \mbox{for $-\pi<\beta<\pi$} \\ 0 & \mbox{else}
               \end{array}\right\}=\delta_{b0}\,;
\nonumber\\  
  \mbox{(b)}\ & \lambda(\alpha)=\delta_{a0}\,,
      \quad     \mu(\beta)=\sinc\left(\frac{\beta}{2}\right)\,;
\\  
  \mbox{(c)}\ & \lambda(\alpha)=\sinc\left(\frac{\alpha+2\pi a}{4}\right)\,,
      \quad     \mu(\beta)=\sinc\left(\frac{\beta+2\pi b}{4}\right)\,;
\nonumber\\ \nonumber
              & \mbox{where $a,b$ are the integers}
                          \quad a=\clst{\frac{\alpha\vphantom{\beta}}{2\pi}}
           \quad\mbox{and}\quad b=\clst{\frac{\beta}{2\pi}}\,.
\end{eqnarray}
We emphasize the symmetry of choice (c), where $\lambda(\gamma)=\mu(\gamma)$
is its own Fourier transform. 
Figure~\ref{fig:mulam-c} shows a plot of this function for $-8\pi<\gamma<8\pi$.

\section{Wave functions: Discrete Zak basis}\label{sec:discreteWFs}
The wave functions for the discrete Zak basis states $\ket{l,m}$ are obtained
from the wave functions in \eqref{cont-wf} by Fourier analysis.
The position wave function
\begin{eqnarray}
  \label{eq:disc-wfx}
  \braket{x}{l,m}&=\int_{(2\pi)}\frac{\D\alpha}{2\pi}
                   \int_{(2\pi)}\frac{\D\beta}{2\pi}\,
                    \braket{x}{\alpha,\beta}\,\Exp{\rmi(l\alpha-m\beta)}
\nonumber\\
                 &=\frac{1}{\sqrt{x_0}}
                   \Exp{\rmi lp_0x/\hbar}\lambda(p_0x/\hbar-2\pi m)
\end{eqnarray}
reveals the physical significance of $\lambda(\alpha)$ of equation
\eqref{mulam2chi}, and we recognize now that \eqref{orthomu} and
\eqref{completemu} are the position representations of the orthogonality and
completeness relations in \eqref{discZak}.

The corresponding momentum wave function
\begin{equation}
  \label{eq:disc-wfp}
  \braket{p}{l,m}=\frac{1}{\sqrt{p_0}}
                   \Exp{-\rmi mx_0p/\hbar}\mu(x_0p/\hbar-2\pi l)
\end{equation}
involves $\mu(\beta)$, so that the Fourier relations of \eqref{mulam} are
actually those between the position and momentum representations.
We note the consistency with \eqref{ladderUV}, which is quite explicitly
visible in
\begin{eqnarray}
  \label{eq:00tolm}
  \ket{l,m}=U^{-l}V^{-m}\ket{0,0}:\ 
&\braket{x}{l,m}=\Exp{\rmi lp_0x/\hbar}\braket{x-mx_0}{0,0}\,,
\nonumber\\
&\braket{p}{l,m}=\Exp{-\rmi mx_0p/\hbar}\braket{p-lp_0}{0,0}\,,
\end{eqnarray}
with
\begin{equation}
  \label{eq:00}
  \braket{x}{0,0}=\lambda(p_0x/\hbar)\big/\sqrt{x_0}\,,\qquad
 \braket{p}{0,0}=\mu(x_0p/\hbar)\big/\sqrt{p_0}\,.
\end{equation}

\begin{figure}[p]
\centerline{%
\begin{picture}(365,460)(0,0)
\put(13,0){\includegraphics{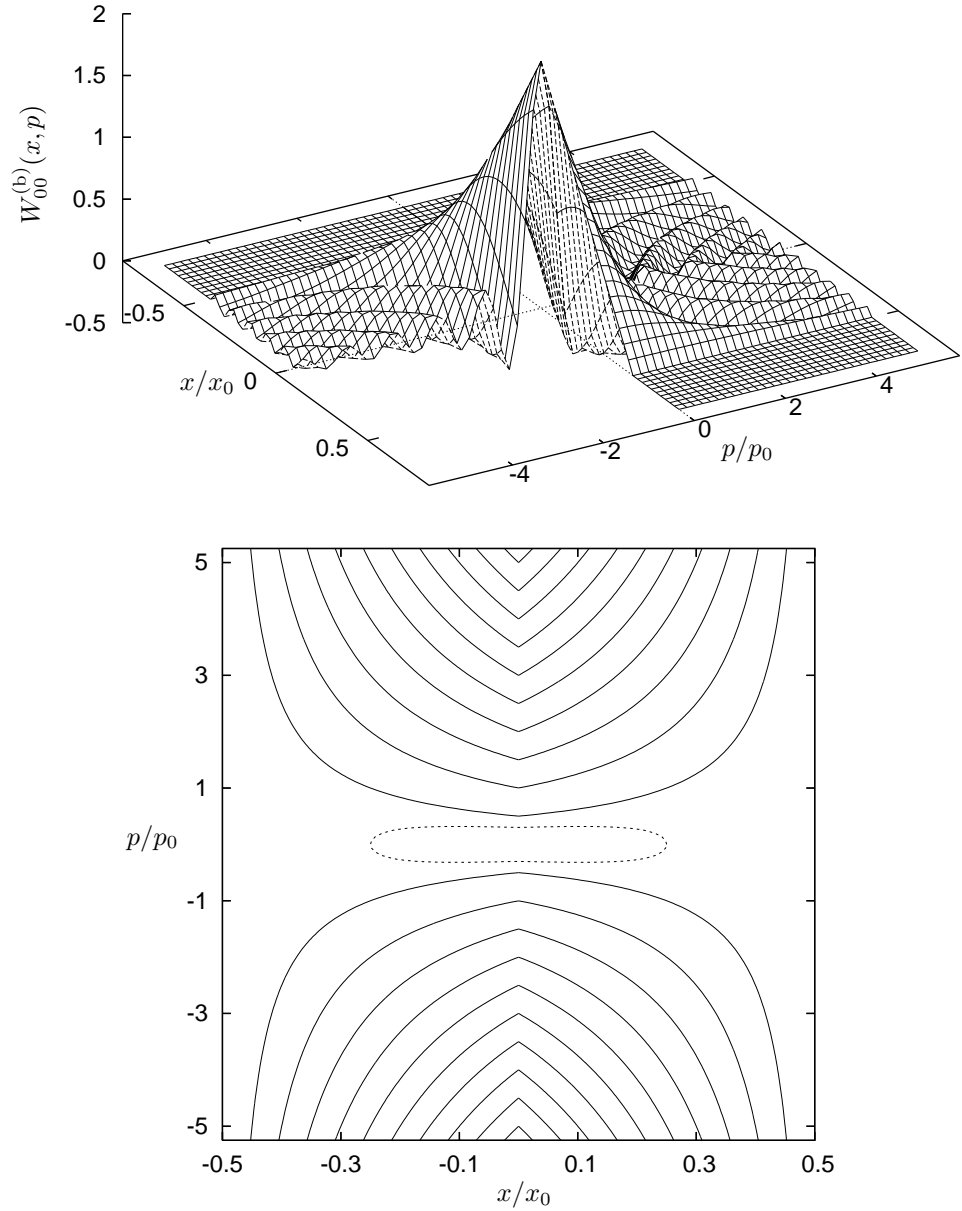}}
\end{picture}
}
\caption{\label{fig:bWig}%
The Wigner function $W^{\mathrm{(b)}}_{00}(x,p)$ corresponding
to the ket $\ket{0,0}$ for choice (b) in equation \eqref{3mulam}.
Top: $W^{\mathrm{(b)}}_{00}$ above the $x,p$ plane, with the quadrant for
$x>0,p<0$ cut out.
Bottom: The $W^{\mathrm{(b)}}_{00}=0$ contour lines where the Wigner function
changes sign (solid lines), and the  $W^{\mathrm{(b)}}_{00}=1$ contour line
where the Wigner function is at half-maximum value (dashed line).} 
\end{figure}

For a visualization of the states of the discrete Zak basis, it may be helpful
to plot their Wigner functions \cite{Wigner1}%
\footnote{\label{fn:Wigner}They are normalized in accordance with 
$\ds\int\frac{\D x\,\D p}{2\pi\hbar}\,W(x,p)=1$ and bounded by
$-2\leq W(x,p)\leq2$.}, given by
\begin{eqnarray}
  \label{eq:lmWigner}
  W_{lm}(x,p)=W_{00}(x-mx_0,p-lp_0)\,,\nonumber\\
  W_{00}(x,p)=2\int\!\frac{\D\alpha}{2\pi}\,\lambda(p_0x/\hbar+\alpha)^*
         \,\Exp{\rmi 2\alpha p/p_0}\, \lambda(p_0x/\hbar-\alpha)\nonumber\\
\phantom{W_{00}(x,p)}=2\int\!\frac{\D\beta}{2\pi}\,\mu(x_0p/\hbar-\beta)^*
         \,\Exp{\rmi 2\beta x/x_0}\, \mu(x_0p/\hbar+\beta)\,,
\end{eqnarray}
so that $W_{lm}$ is centered at $(x,p)=(mx_0,lp_0)$ if $W_{00}$ is centered
at $(x,p)=(0,0)$, as is the case for the three choices in 
\eqref{threeChoices} or \eqref{3mulam}.
Choice (b) yields
\begin{equation}
  \label{eq:bWig00}
  W^{\mathrm{(b)}}_{00}(x,p)=\left\{
    \begin{array}{cl}\displaystyle
    2\snc\left(1-2|x/x_0|,2\pi p/p_0\right) 
    & \mbox{for $|x|<x_0/2$}\,, \\[2ex]  0 & \mbox{for $|x|>x_0/2$}\,,
    \end{array}\right.
\end{equation}
where
\begin{equation}
  \label{eq:snc}
  \snc(z,\alpha)=\frac{\sin(z\alpha)}{\alpha}
=\frac{1}{2}\int\power{z}\rewop{-z}\D\zeta\,\Exp{\rmi\zeta\alpha}\qquad
\mbox{for $0\leq z\leq1$}
\end{equation}
is the incomplete $\sinc$ function, $\sinc(\alpha)=\snc(1,\alpha)$.
For choice (a) the roles of $x$ and $p$ are interchanged, i.e.,
\begin{equation}
  \label{eq:abWig}
   W^{\mathrm{(a)}}_{00}(x,p)=W^{\mathrm{(b)}}_{00}(x_0p/p_0,p_0x/x_0)\,.  
\end{equation}
Figure~\ref{fig:bWig} shows $W^{\mathrm{(b)}}_{00}(x,p)$ for 
$-0.75<x/x_0<0.75$ and $-5.25<p/p_0<5.25$.

\begin{figure}[p]
\centerline{%
\begin{picture}(365,460)(0,0)
\put(13,0){\includegraphics{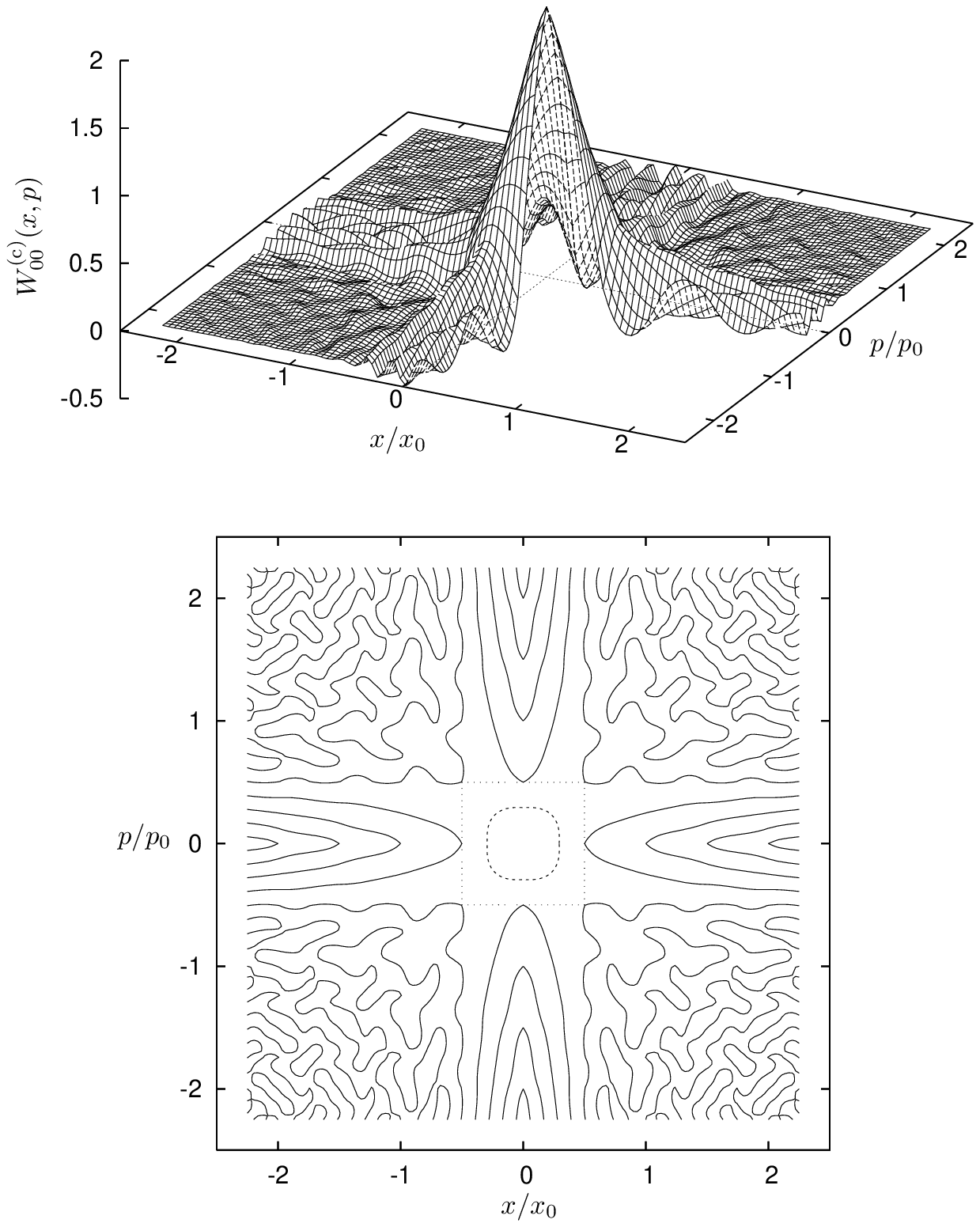}}
\end{picture}
}
\caption{\label{fig:cWig}%
The Wigner function $W^{\mathrm{(c)}}_{00}(x,p)$ corresponding 
to the ket $\ket{0,0}$ for choice (c) in equation \eqref{3mulam}.
Top: $W^{\mathrm{(c)}}_{00}$ above the $x,p$ plane, with the quadrant for
$x>0,p<0$ cut out.
Bottom: The $W^{\mathrm{(c)}}_{00}=0$ contour lines where the Wigner function
changes sign (solid lines), and the  $W^{\mathrm{(c)}}_{00}=1$ contour line
where the Wigner function is at half-maximum value (dashed line).
The dotted square identifies $-\frac{1}{2}<x/x_0,p/p_0<\frac{1}{2}$, the
central peak region of $W^{\mathrm{(c)}}_{00}=0$.} 
\end{figure}

For choice (c) we obtain 
\begin{eqnarray}
  \label{eq:cWig00}
\fl
W^{\mathrm{(c)}}_{00}(x,p)=2\!\!\sum_{j,k=0,1}\!\!(-1)\power{(a+j)(b+k)}
&\snc\left(\bigl|1-|t|-k\bigr|,
           \bigl(2a+s+j\sgn(s)\bigr)\frac{\pi}{2}\right)
\nonumber\\[-2ex]
&\makebox[0pt][r]{$\times$}
\snc\left(\bigl|1-|s|-j\bigr|,
          \bigl(2b+t+k\sgn(t)\bigr)\frac{\pi}{2}\right)\,,
\end{eqnarray}
where $a=\clst{2x/x_0}$ and $b=\clst{2p/p_0}$ are the integers closest to 
$2x/x_0$ and $2p/p_0$, respectively, and $s,t$ account for the differences,
\begin{equation}
  \label{eq:xp2ab}
  2\frac{x}{x_0}=a+s\,,\qquad 2\frac{p}{p_0}=b+t\qquad
\mbox{with}\quad -\frac{1}{2}\leq s,t\leq\frac{1}{2}\,.
\end{equation}
The interchange $p_0x\leftrightarrow x_0p$ has no effect on $
W^{\mathrm{(c)}}_{00}$, 
\begin{equation}
  \label{eq:ccWig}
   W^{\mathrm{(c)}}_{00}(x,p)=W^{\mathrm{(c)}}_{00}(x_0p/p_0,p_0x/x_0)\,,  
\end{equation}
which is yet another manifestation of the symmetry that characterizes the
phase convention (c) of \eqref{threeChoices}.

In effect, then, the phase space is tiled by squares of size 
$\frac{1}{4}x_0p_0=\frac{1}{2}\pi\hbar$, 
centered at $(x,p)=(\half ax_0,\half bp_0)$,
with the pair $s,t$ parameterizing the points of each square relative to the 
center. 
At the centers of the tiles we have
\begin{equation}
  \label{eq:tilecntr}
  W^{\mathrm{(c)}}_{00}(\thalf ax_0,\thalf bp_0)=2\delta_{a0}\delta_{b0}=
\left\{\begin{array}{cl}
2&\mbox{if $a=0$ and $b=0$}\,,\\ 0&\mbox{if $a\neq0$ or $b\neq0$}\,.
\end{array}\right.
\end{equation}
Figure~\ref{fig:cWig} shows $W^{\mathrm{(c)}}_{00}(x,p)$ for 
$-2.25<x/x_0,p/p_0<2.25$, the vicinity of the
maximum at $(x,p)=(0,0)$. 

These plots demonstrate that the Wigner functions in \eqref{bWig00} and
\eqref{cWig00} are continuous in both $x$ and $p$, although their marginals
---  of which
\begin{eqnarray}
  \label{eq:margWig}
  \int\D p\,W(x,p)
      =p_0\bigl|\lambda(p_0x/\hbar)\bigr|^2\,,\nonumber\\
  \int\D x\,W(x,p)
      =x_0\bigl|\mu(x_0p/\hbar)\bigr|^2
\end{eqnarray}
are special cases --- exhibit discontinuities as a rule.
Indeed, it is a matter of inspection to verify that $W^{\mathrm{(b)}}_{00}$
and $W^{\mathrm{(c)}}_{00}$ are continuous.

\section{Operator relations}\label{sec:OperRels}
We combine the wave functions of \eqref{disc-wfx} and \eqref{disc-wfp} with
the Fourier series of \eqref{mulam2chi} to establish
\begin{eqnarray}
  \label{eq:unitLMxp}
\fl  \frac{\bra{x}\Exp{\rmi\alpha L-\rmi\beta M}\ket{p}}{\braket{x}{p}}
&=\sqrt{x_0p_0\,}\,\Exp{-\rmi xp/\hbar}
\sum_{l,m}\braket{x}{l,m}\Exp{\rmi(l\alpha-m\beta)}\braket{l,m}{p}
\nonumber\\
&=\chi\left(\frac{p_0x}{\hbar}+\alpha,\frac{x_0p}{\hbar}\right)^*\!
\EXP{\rmi\frac{\alpha x_0p-\beta p_0x}{4\pi\hbar}}
\chi\left(\frac{p_0x}{\hbar},\frac{x_0p}{\hbar}-\beta\right).
\end{eqnarray}
The normalized matrix elements of $L$ and $M$ are then obtained by
differentiation,
\begin{eqnarray}
  \label{eq:xLMp}
  \frac{\bra{x}L\ket{p}}{\braket{x}{p}}=
  \frac{p}{2p_0}-\chi(\alpha,\beta)^*\frac{1}{\rmi}
      \frac{\partial}{\partial\alpha}\chi(\alpha,\beta)
\biggr|\rewop{\begin{array}[b]{@{}l}%
\alpha=p_0x/\hbar\\\beta=x_0p/\hbar\end{array}}\,,
\nonumber\\
  \frac{\bra{x}M\ket{p}}{\braket{x}{p}}=
  \frac{x}{2x_0}-\chi(\alpha,\beta)^*\rmi
      \frac{\partial}{\partial\beta}\chi(\alpha,\beta)
\biggr|\rewop{\begin{array}[b]{@{}l}%
\alpha=p_0x/\hbar\\\beta=x_0p/\hbar\end{array}}\,.
\end{eqnarray}
The required derivatives are available in \eqref{diffChi}, and so we arrive at
\begin{eqnarray}
  \label{eq:LM}
  \makebox[0pt][r]{(a)\ }
L=\clst{\frac{P}{p_0}}\,,\qquad
M=\frac{X}{x_0}-
\delta^{(2\pi)}\left(\frac{x_0P}{\hbar}-\pi\right)\,
\sin\frac{p_0X}{\hbar}\,,\nonumber\\[1ex]
  \makebox[0pt][r]{(b)\ }
L=\frac{P}{p_0}-
\delta^{(2\pi)}\left(\frac{p_0X}{\hbar}-\pi\right)\,
\sin\frac{x_0P}{\hbar}\,,\qquad
M=\clst{\frac{X}{x_0}}\,,\\[1ex]  \nonumber
  \makebox[0pt][r]{(c)\ }
L=\half\left(\frac{P}{p_0}+\clst{\frac{P}{p_0}}\right)
-\delta^{(2\pi)}\left(\frac{p_0X}{\hbar}-\pi\right)
(-1)\power{\clst{P/p_0}}\sin\frac{x_0P}{2\hbar}\,,
\\\nonumber
M=\half\left(\frac{X}{x_0}+\clst{\frac{X}{x_0}}\right)
-\delta^{(2\pi)}\left(\frac{x_0P}{\hbar}-\pi\right)
(-1)\power{\clst{X/x_0}}\sin\frac{p_0X}{2\hbar}\,,
\end{eqnarray}
which express $L$ and $M$ in terms of $X$ and $P$ for the three phase
conventions of \eqref{threeChoices}.

The reverse relations, which state the line variables $X$ and $P$ in terms of
the torus variables $U,L$, and $V,M$, are at hand as soon as one
observes  that $L-P/p_0$ and $M-X/x_0$ are $2\pi$ periodic functions of
$p_0X/\hbar$ and $x_0P/\hbar$, that is: they are functions of $U$ and $V$. 
For the three choices of \eqref{threeChoices} we have
\begin{eqnarray}
  \label{eq:XP}
\fl  \frac{X}{x_0}=M+\left\{\begin{array}{@{}l}
\mbox{(a)}\
\ds\delta^{(2\pi)}(\beta-\pi)\sin\alpha\,,\\[2ex]
\mbox{(b)}\ 
\ds\frac{\alpha}{2\pi}-\clst{\frac{\alpha}{2\pi}}\,,\\[2ex]
\mbox{(c)}\ 
\ds\half\left(\frac{\alpha}{2\pi}-\clst{\frac{\alpha}{2\pi}}\right)
+\delta^{(2\pi)}(\beta-\pi)(-1)\power{\clst{\frac{\alpha}{2\pi}}}
\sin\frac{\alpha}{2}\,;
  \end{array}\right.
\nonumber \\[-1ex] \qquad\\[-1ex] \nonumber
\fl  \frac{P}{p_0}=L+\left\{\begin{array}{@{}l}
\mbox{(a)}\ 
\ds\frac{\beta}{2\pi}-\clst{\frac{\beta}{2\pi}}\,,\\[2ex]
\mbox{(b)}\ 
\ds\delta^{(2\pi)}(\alpha-\pi)\sin\beta\,,\\[2ex]
\mbox{(c)}\ 
\ds\half\left(\frac{\beta}{2\pi}-\clst{\frac{\beta}{2\pi}}\right)
+\delta^{(2\pi)}(\alpha-\pi)(-1)\power{\clst{\frac{\beta}{2\pi}}}
\sin\frac{\beta}{2}\,;
  \end{array}\right.
\end{eqnarray}
where $\Exp{-\rmi\alpha}=U$ and $\Exp{\rmi\beta}=V$ are understood in the
doubly periodic functions on the right, as illustrated by
\begin{eqnarray}
  \label{eq:halfsine}
  (-1)\power{\clst{\frac{\alpha}{2\pi}}}\sin\frac{\alpha}{2}
&=\sin\left(\frac{\alpha}{2}-\pi\clst{\frac{\alpha}{2\pi}}\right)
=\sum_{j=-\infty}^\infty\frac{(-1)^{j+1}}{(j+\half)\pi}\sin(j\alpha)
\nonumber\\
&
\longrightarrow
\sum_{j=-\infty}^\infty\frac{(-1)^j}{\rmi(2j+1)\pi}\left(U^j-U^{-j}\right)\,.
\end{eqnarray}
For the example of choice (a), then, we have
\begin{eqnarray}
  \label{eq:ZakXP}
  \bra{\alpha,\beta}X=x_0\left(
             -\frac{1}{\rmi}\frac{\partial}{\partial\beta}\bra{\alpha,\beta}
             +\sin\alpha\bra{\alpha,\pi}\right)\,,
\nonumber\\
  \bra{\alpha,\beta}P=p_0\left(\frac{1}{\rmi}\frac{\partial}{\partial\alpha}
                         +\frac{\beta}{2\pi}-\clst{\frac{\beta}{2\pi}}
                         \right)\bra{\alpha,\beta}
\end{eqnarray}
for the differential-operator representation of $X$ and $P$ in the continuous
Zak representation.

\section{Aharonov's modular operators}\label{sec:ModOps}
Adopted to the present notational conventions, Aharonov's modular position and
momentum operators \cite{Aharonov}, $X_\mathrm{m}$ and $P_\mathrm{m}$, are
introduced by the symmetric relations
\begin{equation}
  \label{eq:ModOps}
  X=x_0N_x+X_\mathrm{m}\,,\qquad P=p_0N_p+P_\mathrm{m}\,,
\end{equation}
where $N_x$ and $N_p$ are the integer operators 
\begin{equation}
  \label{eq:intOps}
  N_x=\clst{X/x_0}\,,\qquad N_p=\clst{P/p_0}\,.
\end{equation}
In view of
\begin{equation}
  \label{eq:modUV}
  U=\Exp{-\rmi p_0X_\mathrm{m}/\hbar}\,,\qquad
  V=\Exp{\rmi x_0P_\mathrm{m}/\hbar}\,,
\end{equation}
we can regard the modular operators as the logarithms of the unitary operators
$U$ and $V$ in the sense of
\begin{eqnarray}
  \label{eq:logUV}
  X_\mathrm{m}=X-x_0\clst{X/x_0}
              =\frac{\rmi\hbar}{p_0}
               \ln\frac{\Exp{\epsilon}+U}{\Exp{\epsilon}+U^{-1}}
               \biggr|\rewop{0>\epsilon\to0}\,,\nonumber\\
  P_\mathrm{m}=P-p_0\clst{P/p_0}
              =\frac{\hbar}{\rmi p_0}
               \ln\frac{\Exp{\epsilon}+V}{\Exp{\epsilon}+V^{-1}}
               \biggr|\rewop{0>\epsilon\to0}\,,
\end{eqnarray}
where the limiting procedure ensures ${-\half x_0<X_\mathrm{m}<\half x_0}$
and ${-\half p_0<P_\mathrm{m}<\half p_0}$.

Therefore, $(X_\mathrm{m},N_p)$ is a pair of complementary observables that is
equivalent to $(U,L)$; and, likewise, the pair $(P_\mathrm{m},N_x)$ is as
good as $(V,M)$.  
But the two pairs $(X_\mathrm{m},N_p)$ and $(P_\mathrm{m},N_x)$ together do
not refer to the two \emph{independent} periodic variables of the torus
because $N_p$ does not commute with $N_x$,
\begin{equation}
  \label{eq:NotCommute}
  \rmi\bigl[N_x,N_p\bigr]=\frac{1}{2\pi}-\delta^{(2\pi)}(p_0X/\hbar-\pi)
                          -\delta^{(2\pi)}(x_0P/\hbar-\pi)\,.
\end{equation}
This non-commutativity is as it should be, 
because the pair $(X_\mathrm{m},N_p)$ belongs to the phase convention (a) 
with $N_p=L^{\mathrm{(a)}}$, whereas the pair $(P_\mathrm{m},N_x)$ goes 
with phase convention (b), as $N_x=M^\mathrm{(b)}$. 

The non-commutativity of $N_x$ and $N_p$ reminds us that it is not possible to
constrain a quantum system both to a finite $x$ range and to a finite $p$ 
range. 
It appears that, in the present context of the Zak bases, the symmetric
compromise of imperfect simultaneous localization is given by the discrete Zak
basis states of the symmetric phase convention (c). 
Indeed, $W^{\mathrm{(c)}}_{00}(x,p)$ is very strongly peaked within the
central square $-\half\lesssim x/x_0,p/p_0\lesssim\half$ that is indicated in
Figure \ref{fig:cWig}.

\section{Hinting at an application: Toroidal qubits}\label{sec:TorQbits}
The three operators that are introduced by
\begin{equation}
  \label{eq:sigDef}
  \sigma_1+\rmi\sigma_2=\bigl[1+(-1)^L\bigr]U\,,\qquad \sigma_3=(-1)^L
\end{equation}
are such that
\begin{eqnarray}
  \label{eq:sigmas}
  \trc{\sigma_1}=\trc{\sigma_2}=\trc{\sigma_3}=0\,,\nonumber\\
  \sigma_1^2=\sigma_2^2=\sigma_3^2=1\,,\nonumber\\
  \sigma_1\sigma_2=\rmi\sigma_3\,,\quad
  \sigma_2\sigma_3=\rmi\sigma_1\,,\quad
  \sigma_3\sigma_1=\rmi\sigma_2\,,
\end{eqnarray}
so that they are the components of a genuine Pauli vector operator 
$\vec{\sigma}=(\sigma_1,\sigma_2,\sigma_3)$.
We can, therefore, use them to associate a qubit with the $U,L$  
degree of freedom.
Likewise,
\begin{equation}
  \label{eq:tauDef}
    \tau_1+\rmi\tau_2=\bigl[1+(-1)^M\bigr]V\,,\qquad \tau_3=(-1)^M
\end{equation}
defines a second toroidal qubit for the $V,M$ degree of freedom.

For any quantum state on the line, there is then a two-qubit state with the
statistical operator
\begin{equation}
  \label{eq:rho2qb}
  \rho=\frac{1}{4}\bigl(1+\vec{\sigma}\cdot\expect{\vec{\sigma}}
                         +\expect{\vec{\tau}}\cdot\vec{\tau}
 +\vec{\sigma}\cdot\expect{\vec{\sigma}\,\vec{\tau}}\cdot\vec{\tau}\,\bigr)\,.
\end{equation}
The two vectorial expectation values $\expect{\vec{\sigma}}$, 
$\expect{\vec{\tau}}$ together with the dyadic expectation value 
$\expect{\vec{\sigma}\,\vec{\tau}}$ constitute the 15 parameters that specify
the two-qubit state.
Once their values are known, one can, for instance, decide whether $\rho$ is
separable or not, and thus whether an entangled qubit pair is concealed in the
given state on the line.
This application and others are, however, beyond the scope of the present
paper on the properties of the Zak bases.

\section{Summary}
The unitary operators for displacement in position and momentum
commute with one another when the product of the elementary
displacements forms the elementary area $h=2\pi\hbar$ in phase space.
The Zak basis is composed of the complete set of eigenstates of 
these commuting operators and is therefore specified by two phase 
parameters.

Hence the Zak basis maps the Hilbert space on the line onto 
the Hilbert space on a torus. 
We showed that for this mapping to be one to one, periodicity of the Zak basis
states in their phase parameters is required and this, in turn, imposed
discontinuity in the phases.  
We used the freedom present for the definition of this phases discontinuity to
consider three choices which we deemed most natural: 
(i) one phase parameter continuous, say $\alpha$, and the other, 
$\beta$, discontinuous; 
(ii) conversely, $\alpha$ discontinuous while $\beta$ continuous; 
and (iii) symmetric discontinuity in both phases. 
We thereby generalized Zak's approach, whose original choice
\cite{Zak1,Zak2,Zak3,Zak4} was asymmetric.  

The resultant periodic Zak basis may be represented by a double Fourier 
series. 
The Fourier coefficients constitute a doubly discrete representation basis
that is mutually unbiased with the Zak basis. 
The states of this discrete representation are characterized by pairs of
integer eigenvalues.
The hermitian operators, for which these states are eigenstates with these
eigenvalues, generate the fundamental rotations of the torus. 
Thus with each consistent phase mapping convention of the line onto the
torus we have one pair of mutually unbiased Zak bases.

Further, we considered briefly the relation between the Zak bases and
Aharonov's modular operators and interpreted the operators employed
by the latter in terms of the former: 
both, per force, reflect the basic fact that it is not possible to constrain a
quantum system to finite domains in \emph{both} position and momentum.

As a possible application of our study of the Zak operators on the
torus we mentioned the possibility of associating with a single  
degree of freedom on the line a pair of potentially entangled
qubits.
This is achieved by a rather simple, and quite natural, construction of two
sets of Pauli operators in terms of the basic unitary operators of the Zak
bases.

\section*{Acknowledgments}
We are grateful for helpful discussions with Joshua Zak and Yakir Aharonov.
AM and MR acknowledge the very kind hospitality extended to them by 
the National University of Singapore, and in particular by 
Prof B-G~Englert.
BGE wishes to thank Prof C Miniatura for the generous hospitality at the
Institut Nonlin\'eaire de Nice where part of this work was done.
This work was supported in part by the Singapore A*STAR Temasek Grant
No.~012-104-0040.

\end{document}